\documentstyle[a4,12pt,axodraw]{article}
\newcommand{\plus}{\makebox[15pt][c]{$+$}}
\newcommand{\minus}{\makebox[15pt][c]{$-$}}
\newcommand{\err}[2]{
\raisebox{0.08em}{\scriptsize {$\hspace{-0.3em}\begin{array}{@{}l@{}}
                          \plus\makebox[0.55em][r]{#1} \\[-0.12em]
                          \minus\makebox[0.55em][r]{#2}
                        \end{array}$}}}
\newcommand{\er}[2]{
\raisebox{0.08em}{\scriptsize {$\hspace{-0.3em}\begin{array}{@{}l@{}}
                          \plus\makebox[0.15em][r]{#1} \\[-0.12em]
                          \minus\makebox[0.15em][r]{#2}
                        \end{array}$}}}
\newcommand{\errr}[2]{
\raisebox{0.08em}{\scriptsize {$\hspace{-0.3em}\begin{array}{@{}l@{}}
                          \plus\makebox[0.9em][r]{#1} \\[-0.12em]
                          \minus\makebox[0.9em][r]{#2}
                        \end{array}$}}}

\newcommand{\ewxy}[2]{\setlength{\epsfxsize}{#2}\epsfbox[10 30 640
  590]{#1}}

\begin{document}
\begin{titlepage}

\begin{flushright}
CERN-TH/96-257\\ 
SHEP 96-23
\end{flushright}
\vspace{1in}

\large
\centerline {\bf Exclusive Decays of Beauty Hadrons}
\normalsize
 
\vskip 2.0cm \centerline {C.T. Sachrajda~\footnote{Invited lecture, presented
at the workshop ``Beauty `96'', Rome, 17-21 June 1996.}}
\centerline {\it Theory Division, CERN, CH-1211 Geneva 23 Switzerland}
\centerline{\it and} 
\centerline{\it Physics Department, University of
Southampton, Southampton SO17 1BJ, UK} \vskip 4.0cm
 
\centerline {\bf Abstract} \vskip 1.0cm 

The principal difficulty in deducing weak interaction properties from
experimental measurements of $B$-decays lies in controlling the strong
interaction effects. In this talk I review the status of theoretical
calculations of the amplitudes for exclusive leptonic and semileptonic
decays, in the latter case with special emphasis on the
extraction of the $V_{cb}$ and $V_{ub}$ matrix elements.  \vfill

\end{titlepage}

\newpage
\section{Introduction}
\label{sec:intro}

In this lecture I will review the status of theoretical calculations of
exclusive $B$-decays. It is intended that this talk should complement
those presented at this conference by N.~Uraltsev~\cite{kolya}
(theory of heavy quark physics), A.~Ali~\cite{ahmed} (rare
$B$-decays) and M.~Gronau~\cite{gronau} ($CP$-violation). 
The two main topics which will be discussed here are: 
\begin{itemize}
\item[i)] {\em Leptonic Decays} in which the $B$-meson decays into
  leptons, e.g. $B\to \tau\nu_\tau$. These are the simplest
  to consider theoretically (see sec.~\ref{sec:fb}). Their observation
  at future $b$-factories would have a significant impact on the
  phenomenology of beauty decays.
\item[ii)] {\em Semileptonic Decays} in which the $b$-quark decays
  into a lighter quark + leptons.  Examples of such decays include
  $B\to (D\ {\mathrm or}\ D^*) + l\nu_l$ and $B\to (\pi\ {\mathrm or}\ 
  \rho) + l\nu_l$, which are being used to determine the $V_{cb}$ and
  $V_{ub}$ matrix elements of the CKM-matrix (see
  sec.~\ref{sec:semilept}). Many of the theoretical issues concerning
  these decays are relevant also for rare decays, such as $B\to
  K^*\gamma$.
\end{itemize}
{\em Non-Leptonic Decays} in which the $B$-meson decays into two or
more hadrons, such as $\bar B^0\to\pi^-D^+$, are considerably more
complicated to treat theoretically, and with our current level of
understanding require model assumptions. I will not discuss them
further in this talk (see however the talk by Gronau~\cite{gronau}).

In studying the decays of $B$-mesons, we are largely interested in
extracting information about the properties and parameters of the weak
interactions, and in looking for possible signatures of physics beyond
the standard model. The most important theoretical problem in
interpreting the experimental results, is to control the strong
interaction effects which are present in these decays. This is a
non-perturbative (and hence very difficult) problem, and is the main
subject of this talk.  The main theoretical tools that are used to
quantify the effects are lattice simulations and QCD sum rules,
combined with the formalism of the heavy quark effective theory (HQET)
where appropriate. 

As with any problem in non-perturbative quantum field theory, the
exploitation of all available symmetries is very important. For the
case of heavy quark physics, the use of the spin-flavour symmetries
that are present when the masses of the heavy quarks are $\gg
\Lambda_{QCD}$, leads to considerable simplifications (see
refs.~\cite{kolya} and \cite{mn,ms} for recent reviews and references
to the original literature). In particular, as will be seen in the
following sections, the use of heavy quark symmetries and the
HQET is particularly helpful for $B$-decays.

It is not appropriate in this lecture to present a detailed critical
review of the systematic errors present in lattice simulations (see
ref.~\cite{lattbphys} for a recent review).  Since many of the results
below are based on lattice simulations, it is, however, necessary to
mention at least the main source of uncertainty present in the
calculations of quantities in $B$-physics. The number of space time
points on a lattice is limited by the available computing resources.
One therefore has to compromise between two competing requirements:
(i) that the lattice be sufficiently large in physical units to
contain the particle(s) whose properties are being studied, i.e. the
length of the lattice in each direction should be $\gg 1$\,fm, and
(ii) that the spacing between neighbouring lattice points, $a$, be
sufficiently small to avoid errors due to the granularity of the
lattice (called ``lattice artefacts'' or ``discretization errors'' in
the literature), i.e.  $a^{-1}\gg \Lambda_{QCD}$. Much effort is
currently being devoted to reducing the discretization effects by
constructing ``improved'' (or even ``perfect''~\cite{perfect}) lattice
actions and operators following the approach of
Symanzik~\cite{improvement}.  Typical values of $a^{-1}$ in current
simulations are about 2--3 GeV, i.e. the lattice spacings are larger
than the Compton wavelength of the $b$-quark, and the propagation of a
$b$-quark on such lattices cannot be studied directly.  The results
presented below are obtained by extrapolating those computed directly
for lighter quarks (with masses typically around that of the charm
quark).  In addition, calculations can be performed in the HQET and
the results obtained in the infinite mass limit can then be used to
guide this extrapolation. I should also add that, except where
explicitly stated to the contrary, the results below have been
obtained in the quenched approximation, in which sea-quark loops are
neglected. This approximation is very gradually being relaxed, as
computing resources and techniques are improved.

The second non-perturbative method which is used extensively to
compute amplitudes for $B$-decays is QCD sum rules~\cite{svz}. In this
approach, correlation functions are calculated at intermediate
distances, keeping a few terms in the Operator Product Expansion
(OPE), and by using dispersion relations are related to spectral
densities. The evaluation of the systematic uncertainties, such as
those due to the truncation of the perturbation series and OPE or to
the specific models that are used for the continuum contribution to
the spectral densities, is a very complicated issue; see
refs.~\cite{mn,ms} and the papers cited below for any discussion of
this important question.

I now review the status of leptonic and semileptonic decays of
$B$-mesons in turn.

\section{Leptonic Decays}
\label{sec:fb}

\begin{figure}[t]
\begin{center}
\begin{picture}(180,40)(0,15)
\SetWidth{2}\ArrowLine(10,41)(50,41)
\SetWidth{0.5}
\Oval(80,41)(15,30)(0)
\ArrowLine(79,56)(81,56)\ArrowLine(81,26)(79,26)
\GCirc(50,41){7}{0.8}
\Photon(110,41)(150,41){3}{7}
\ArrowLine(150,41)(167,51)\ArrowLine(167,31)(150,41)
\Text(20,35)[tl]{${\mathbf B^-}$}
\Text(80,62)[b]{$b$}\Text(80,20)[t]{$\bar u$}
\Text(172,53)[l]{$l^-$}\Text(172,30)[l]{$\bar\nu$}
\Text(132,48)[b]{$W$}
\end{picture}
\caption{Diagram representing the leptonic decay of the $B$-meson.}
\label{fig:fb}
\end{center}
\end{figure}
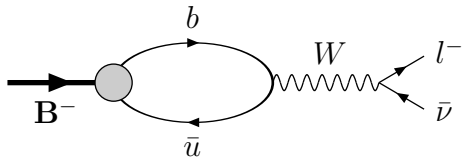

Leptonic decays of $B$-mesons, see fig.~\ref{fig:fb}, are particularly
simple to treat theoretically~\footnote{For simplicity the
  presentation here is for the pseudoscalar $B$-meson. A parallel
  discussion holds also for the vector meson $B^*$.}. The strong
interaction effects are contained in a single \underline{unknown} number,
called the decay constant $f_B$. Parity symmetry implies that only the
axial component of the $V$--$A$ weak current contributes to the decay,
and Lorentz invariance that the matrix element of the axial current is
proportional to the momentum of the $B$-meson (with the constant of
proportionality defined to be $f_B$):
\begin{equation}
\langle 0\,|\, A_\mu(0)\, |\, B(p)\rangle = i\, f_B \, p_\mu\ .
\label{eq:fbdef}\end{equation}
Knowledge of $f_B$ would allow us to predict the rates for the
corresponding decays:
\begin{equation}
\Gamma(B\to\,l\nu_l\, + \,l\nu_l\,\gamma) = 
\frac{G_F^2 V_{ub}^2}{8\pi}f_B^2m_l^2m_B
\left(1 - \frac{m_l^2}{m_B^2}\right)^2\,( 1 + O(\alpha))\ ,
\label{eq:rate}\end{equation}
where the $O(\alpha)$ corrections are also known.

In addition to leptonic decays, it is expected that the knowledge of
$f_B$ would also be useful for our understanding of other processes in
$B$-physics, particularly for those for which ``factorization'' might
be thought to be a useful approximation. For example, in
$B$--$\overline{B}$ mixing, the strong interaction effects are
contained in the matrix element of the $\Delta B$\,=\,2 operator:
\begin{equation}
M = \langle \overline{B}^0\,|\,\bar b \gamma_\mu(1-\gamma_5)q\ 
\bar b \gamma_\mu(1-\gamma_5)q\,|\,B^0\rangle\ .
\label{eq:mbbar}\end{equation}
It is conventional to introduce the $B_B$-parameter through the definition
\begin{equation}
M = \frac{8}{3}\, f_B^2 M_B^2 B_B \ .
\label{eq:bbdef}\end{equation}
In the vacuum saturation approximation (whose precision is difficult
to assess a priori) $B_B=1$. It appears that $B_B$ is considerably
easier to evaluate than $f_B$, e.g. recent lattice results (for the
matrix element $M$ of the operator renormalized at the scale $m_B$ in
the $\overline{MS}$ scheme) include $B(m_b) = 0.90(5)$ and
0.84(6)~\cite{jlqcdbb} and 0.90(3)~\cite{bbs}.  Thus it is likely that
the uncertainty in the value of the matrix element $M$ in
eq.~(\ref{eq:mbbar}) is dominated by our ignorance of $f_B$.

\paragraph {${\mathbf f_{D_s}}$:}
Since experimental results are beginning to become available for
$f_{D_s}$, I will start with a brief review of the decay constants of
charmed mesons. Many lattice computations of $f_D$ have been performed
during the last ten years, and my summary of the results
is~\cite{marseille}~\footnote{\label{foot:fdfb} The rapporteur at the
  1995 Lattice conference summarized the results for the decay
  constants as $f_D\simeq f_B\simeq 200\,\mbox{GeV}\pm 20\%
  $~\protect\cite{allton}.}
\begin{equation}
f_D = 200 \pm 30\ \mbox{MeV}\ ,
\label{eq:fdlatt}\end{equation}
using a normalization in which
$f_{\pi^+}\simeq 131$~MeV.
The value of the decay constant is found to decrease as the mass of
the light valence quark is decreased (as expected), so that $f_{D_s}$
is 7--15\% larger than $f_D$, $f_{D_s}=220\pm 35$~MeV. As an
example of the many lattice results which have been published for
$f_{D_s}$, I give here the two new ones presented at this year's
international symposium on lattice field theory. The MILC
collaboration found $f_{D_s} = 211 \pm 7 \pm 25\pm 11$ MeV, where the
first error is statistical, the second an estimate of the
systematic uncertainties within the quenched approximation, and the
third an estimate of the quenching errors~\cite{milc}. The JLQCD
collaboration found $f_{D_s} = 216\pm 6\err{22}{15}$~MeV, where the
second error is systematic (within the quenched
approximation)~\cite{jlqcdfds}. These results illustrate the fact that
the errors are dominated by systematic uncertainties, and the main
efforts of the lattice community are being devoted to controlling
these uncertainties.

It is very interesting to compare the lattice \underline{prediction}
of $220\pm 35$~MeV with experimental measurements for $f_{D_s}$.  The
1996 Particle Data book~\cite{pdg} quotes the results $f_{D^+}<
310$~MeV and
\begin{eqnarray}
f_{D_s^+} & = & 232 \pm 45 \pm 20 \pm 48 \ \mbox{MeV}\hspace{1in}\mbox{WA75}
\label{eq:fdspdgwa75}\\ 
f_{D_s^+} & = & 344 \pm 37 \pm 52 \pm 42 \ \mbox{MeV}\hspace{1in}\mbox{CLEO}
\label{eq:fdspdgcleo}\\ 
f_{D_s^+} & = & 430 \errr{150}{130} \pm 40  \ \mbox{MeV}\hspace{1.5in}
\label{eq:fdspdgbes}\mbox{BES}\ .
\end{eqnarray}
More recently the CLEO result has been updated~\cite{cleoupdate}
($f_{D_s^+}= 284 \pm 30 \pm 30 \pm 16$~MeV) and the E653 collaboration
has found~\cite{e653} $f_{D_s^+}= 194 \pm 35 \pm 20 \pm 14$~MeV.
Combining the four measurements of $f_{D_s}$ from $D_s\to\mu\nu$
decays, the rapporteur at this year's ICHEP conference
found~\cite{richman}
\begin{equation}
f_{D_s} = 241 \pm 21 \pm 30\ \mbox{MeV}\ .
\label{eq:richman}\end{equation}
In spite of the sizeable errors, the agreement with the lattice
prediction is very pleasing and gives us further confidence in the
predictions for $f_B$ and related quatnities.

\paragraph {${\mathbf f_B}$:}  
For sufficiently large masses of the heavy quark, the decay constant
of a heavy--light pseudoscalar meson ($P$) scales with its mass ($M_P$) as
follows:
\begin{equation}
f_P = \frac{A}{\sqrt{M_P}}\left[\alpha_s(M_P)^{-2/\beta_0}\left\{1 + 
O(\alpha_s(M_P)\,)\,\right\} + O\left(\frac{1}{M_P}\right)\,\right]\ ,
\label{eq:fpscaling}\end{equation}
where $A$ is independent of $M_P$.
Using the scaling law~(\ref{eq:fpscaling}), a value of about 200~MeV
for $f_D$ would correspond to $f_B\simeq 120$~MeV. Results from
lattice computations, however, indicate that $f_B$ is significantly
larger than this and that the $O(1/M_P)$ corrections on the right-hand
side of eq.~(\ref{eq:fpscaling}) are considerable. My summary of the
lattice results is~\cite{marseille} (see also footnote~\ref{foot:fdfb}):
\begin{equation}
f_B = 180 \pm 40\ \mbox{MeV}\ .
\label{eq:fblatt}\end{equation}
The coefficient of the $O(1/M_P)$ corrections is found to be typically 
between 0.5 and 1~GeV.

Present lattice studies of heavy--light decay constants are
concentrating on relaxing the quenched approximation, on calculating
the $O(1/M_P)$ corrections in eq.~(\ref{eq:fpscaling}) explicitly, and
on reducing the discretization errors through the use of improved
actions and operators. The results obtained using QCD sum rules are
in very good agreement with those from lattice simulations (see, for
instance, ref.~\cite{mn} and references therein, and
ref.~\cite{narisoncracow}).

\section{Semileptonic Decays}
\label{sec:semilept}

\begin{figure}
\begin{center}
\begin{picture}(180,60)(20,10)
\SetWidth{2}\ArrowLine(10,41)(43,41)
\Text(15,35)[tl]{${\mathbf B}$}
\SetWidth{0.5}
\Oval(100,41)(20,50)(0)
\SetWidth{2}\ArrowLine(157,41)(190,41)
\Text(180,35)[tl]{${\mathbf D,\,D^*,\,\pi,\,\rho}$}
\SetWidth{0.5}
\Vertex(100,61){3}
\GCirc(50,41){7}{0.8}\GCirc(150,41){7}{0.8}
\Text(75,48)[b]{$b$}\Text(117,48)[b]{$c,u$}
\Text(100,16)[t]{$\bar q$}
\Text(100,71)[b]{$V$--$A$}
\ArrowLine(101,21)(99,21)
\ArrowLine(70,57)(72,58)\ArrowLine(128,57.5)(130,56.5)
\end{picture}
\caption{Diagram representing the semileptonic decay of the $B$-meson.
$\bar q$ represents the light valence antiquark, and the black circle
represents the insertion of the $V$--$A$ current with the appropriate
flavour quantum numbers.}
\label{fig:sl}
\end{center}
\end{figure}
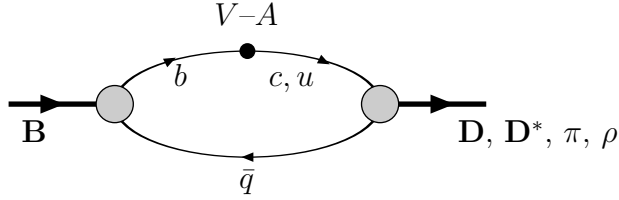

For the remainder of this talk I will discuss semileptonic decays of
$B$-mesons, considering in turn the two cases in which the $b$-quark
decays semileptonic\-ally into a $c$-quark or a $u$-quark, see
fig.~\ref{fig:sl}. In both cases it is convenient to use space-time
symmetries to express the matrix elements in terms of invariant
form factors (I use the helicity basis for these as defined below).
When the final state is a pseudoscalar meson $P=D$ or $\pi$, parity
implies that only the vector component of the $V$--$A$ weak current
contributes to the decay, and there are two independent form factors,
$f^+$ and $f^0$, defined by
\begin{eqnarray}
\langle P(p_P)| V^\mu|B(p_B)\rangle & = &
f^+(q^2)\left[\,(p_B+p_P)^\mu -
\frac{M_B^2 - M_P^2}{q^2}\,q^\mu\right] \nonumber\\ 
& + & \ \ \ f^0(q^2)\,\frac{M_B^2 - M_P^2}{q^2}\,q^\mu\ ,
\label{eq:ffpdef}\end{eqnarray}
where $q$ is the momentum transfer, $q=p_B-p_P$. When the final-state
hadron is a vector meson $V=D^*$ or $\rho$, there are four independent
form factors:
\begin{eqnarray}
\langle V(p_V)| V^\mu|B(p_B)\rangle & = &
\frac{2V(q^2)}{M_B+M_V}\epsilon^{\mu\gamma\delta\beta}
\varepsilon^*_\beta p_{B\,\gamma}p_{V\,\delta}\label{eq:ffvvdef}\\ 
\langle V(p_V)| A^\mu|B(p_B)\rangle  & = &  
i (M_B + M_V) A_1(q^2) \varepsilon^{*\,\mu}\, - \nonumber\\ 
&&\hspace{-1in}
i\frac{A_2(q^2)}{M_B+M_V} \varepsilon^*\hspace{-3pt}\cdot\hspace{-2pt}p_B 
(p_B + p_V)^\mu + i \frac{A(q^2)}{q^2} 2 M_V 
\varepsilon^*\hspace{-3pt}\cdot\hspace{-2pt}p_B q^\mu\ ,
\label{eq:ffvadef}\end{eqnarray}
where $\varepsilon$ is the polarization vector of the final-state meson,
and $q = p_B-p_V$. 
Below we shall also discuss the form factor $A_0$, which is given 
in terms of those defined above by $A_0 = A + A_3$, with
\begin{equation}
A_3 = \frac{M_B + M_{D^*}}{2 M_{D^*}}A_1 - 
\frac{M_B - M_{D^*}}{2 M_{D^*}}A_2\ .
\label{eq:a3def}\end{equation}

\subsection{Semileptonic ${\mathbf B\to D}$ and ${\mathbf B\to D^*}$ 
Decays}
\label{subsec:vcb}

$B\to D^*$ and, more recently, $B \to D$ decays are used to determine
the $V_{cb}$ element of the CKM matrix. Theoretically they are 
relatively simple to consider, since the heavy quark symmetry
implies that the six form factors are related, and that there is only
one independent form factor $\xi(\omega)$, specifically:
\begin{eqnarray}
f^+(q^2) & = & V(q^2) = A_0(q^2) = A_2(q^2) 
\nonumber\\ 
& = & \left[1 - 
\frac{q^2}{(M_B + M_D)^2}\right]^{-1} A_1(q^2) = \frac{M_B+M_D}
{2\sqrt{M_BM_D}}\,\xi(\omega)\ ,
\label{eq:iw}\end{eqnarray}
where $\omega = v_B\cdot v_D$. Here the label $D$ represents the $D$-
or $D^*$-meson as appropriate. In this leading approximation the
pseudoscalar and vector mesons are degenerate. The unique form factor
$\xi(\omega)$, which contains all the non-perturbative QCD effects, is
called the Isgur--Wise (IW) function. Vector current conservation
implies that the IW-function is normalized at zero recoil, i.e. that
$\xi(1) =1$. This property is particularly important in the extraction
of the $V_{cb}$ matrix element.

The relations in eq.~(\ref{eq:iw}) are valid up to perturbative and
power corrections. The theoretical difficulty in making predictions
for the form factors lies in calculating these corrections with
sufficient precision.

The decay distribution for $B\to D^*$ decays can be written as:
\begin{eqnarray}
\frac{d\Gamma}{d\omega} & = & \frac{G_F^2}{48\pi^3}
(M_B-M_{D^*})^2 M_{D^*}^3 \sqrt{\omega^2 -1}\,(\omega+1)^2\cdot
\nonumber\\ 
& &\hspace{-0.35in}\left[ 1 + \frac{4\omega}{\omega + 1} 
\frac{M_B^2 - 2\omega M_BM_{D^*}+M_{D^*}^2}{(M_B-M_{D^*})^2}\right]
|V_{cb}|^2\, {\cal F}^2(\omega)\ ,
\label{eq:distr}\end{eqnarray}
where ${\cal F}(\omega)$ is the IW-function combined with perturbative
and power corrections. It is convenient theoretically to
consider this distribution near $\omega = 1$. In this case $\xi(1) =
1$, and there are no $O(1/m_Q)$ corrections (where $Q= b$ or $c$) by
virtue of Luke's theorem~\cite{luke}, so that the expansion of ${\cal
  F}(1)$ begins like:
\begin{equation}
{\cal F}(1) = \eta_A\left( 1 + 0\,\frac{\Lambda_{QCD}}{m_Q} + 
c_2\frac{\Lambda^2_{QCD}}{m_Q^2} + \cdots\right)\, ,
\label{eq:fexpansion}\end{equation}
where $\eta_A$ represents the perturbative corrections.
The one-loop contribution to $\eta_A$ has been known for some time now,
whereas the two-loop contribution was evaluated this year, with the
result~\cite{czarnecki}:
\begin{equation}
\eta_A = 0.960\pm 0.007\ ,
\end{equation}
where we have taken the value of the two loop contribution as an
estimate of the error.

The power corrections are much more difficult to estimate reliably.
Neubert has recently combined the results of
refs.~\cite{fn}--\cite{suv} to estimate that the $O(1/m_Q^2)$ terms in
the parentheses in eq.~(\ref{eq:fexpansion}) are about $-0.055\pm
0.025$ and that
\begin{equation}
{\cal F}(1) = 0.91 (3)\ .
\label{eq:f1result}\end{equation}

In considering eq.~(\ref{eq:f1result}), the fundamental question that
has to be asked is whether the power corrections are sufficiently
under control. There are differing, passionately held views on this
subject. The opinion of G.~Martinelli and myself is that the
uncertainty in eq.~(\ref{eq:f1result}) is
underestimated~\cite{ht}.  The power corrections are proportional
to matrix elements of higher-dimensional operators. These have either
to be evaluated non-perturbatively or to be determined from some other
physical process.  In either case, before the matrix element can be
determined a subtraction of large terms is required (since
higher-dimensional operators in general contribute to non-leading
terms).  The ``large'' terms are usually only known in perturbation
theory at tree level, one-loop level or exceptionally at two-loop
level.  Therefore the precision of such a subtraction is limited.
Moreover the definition of the higher-dimensional operators, and hence
the value of their matrix elements, depend significantly on the
treatment of the higher-order terms of the perturbation series for the
coefficient function of the leading twist operator (this series not
only diverges, but is not summable by any standard technique).  These
arguments are expanded, with simple examples and references to the
original literature, in ref.~\cite{ht}. Considerable effort is being
devoted at present to improving the theoretical control over power
corrections.

Bearing in mind the caveat of the previous paragraph, the procedure
for extracting the $V_{cb}$ matrix element is to extrapolate the
experimental results for $d\Gamma/d\omega$ to $\omega = 1$ and to use
eq.~(\ref{eq:distr}) with the theoretical value of ${\cal F}(1)$. See
for example the results presented by Artuso at this
conference~\cite{artuso}.

\begin{figure}
\begin{picture}(120,200)
\put(50,-20){\ewxy{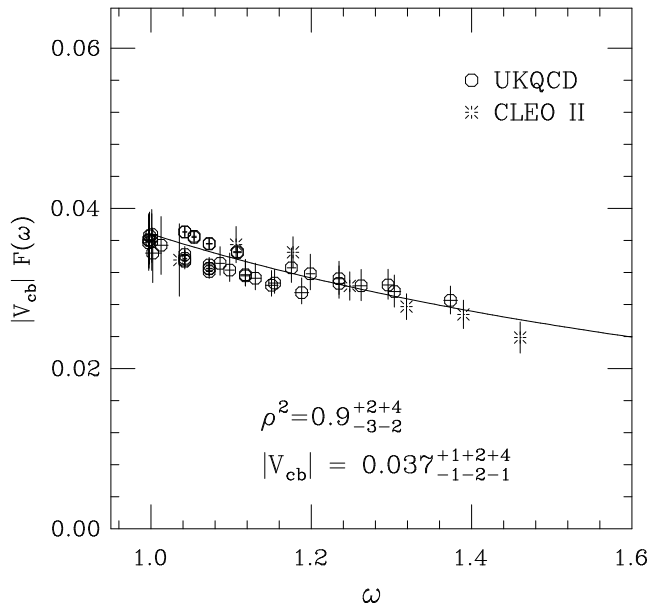}{110mm}} 
\end{picture}
\caption{Fit of the UKQCD lattice results for 
  $|V_{cb}|{\cal F}(\omega)$~\protect\cite{ukqcdiw}
  to the experimental data from the CLEO collaboration
  ~\protect\cite{cleoii}.}
\label{fig:iw}\end{figure}

Having discussed the theoretical status of the normalization ${\cal
  F}(1)$, let us now consider the shape of the function ${\cal
  F}(\omega)$, near $\omega = 1$. A theoretical understanding of the
shape would be useful to guide the extrapolation of the experimental
data, and also as a test of our understanding of the QCD effects. We
expand ${\cal F}$ as a power series in $\omega -1$:
\begin{equation}
{\cal F}(\omega) = {\cal F}(1)\, \left[1 - \hat\rho^2\,(\omega -1)
+\hat c\, (\omega -1 )^2 + \cdots\right]\ ,
\label{eq:ftaylor}\end{equation}
where~\cite{neubertcernschool}
\begin{equation}
\hat\rho^2 = \rho^2 + (0.16\pm 0.02) + \mbox{power corrections}\ ,
\label{eq:rhohat}\end{equation}
and $\rho^2$ is the slope of the IW-function. What is known theoretically
about the parameters in eqs.~(\ref{eq:ftaylor}) and (\ref{eq:rhohat})?  
Bjorken~\cite{bj} and Voloshin~\cite{voloshin} have derived lower and
upper bounds, respectively, for the $\rho^2$:
\begin{equation}
\frac{1}{4} < \rho^2 < \frac{1}{4} + \frac{\overline{\Lambda}}{2 E_{min}}
\ ,
\label{eq:rhosqbounds}\end{equation}
where $\overline\Lambda$ is the binding energy of the $b$-quark in the 
$B$-meson, and $E_{min}$ is the difference in masses between the ground
state and the first excited state.  There are perturbative corrections
to the bounds in eq.~(\ref{eq:rhosqbounds})~\cite{gk}, on the basis of
which Korchemsky and Neubert~\cite{kn} conclude that
\begin{equation}
0.5 < \rho^2 < 0.8\ .
\label{eq:rhosqbounds2}\end{equation} 
Values of $\rho^2$ obtained using QCD sum rules and lattice
simulations are presented in table~\ref{tab:rhosq}. The theoretical
results are broadly in agreement with the experimental measurements,
e.g. in fig.~\ref{fig:iw} we show the comparison of the lattice
results from ref.~\cite{ukqcdiw} with the data from the CLEO
collaboration~\cite{cleoii}.

\begin{table}[htb]
\centering
\begin{tabular}{|c|l|}
\hline
$\rho^2$ & \hspace{0.4in}Method\\ \hline
$0.84\pm 0.02$ & QCD sum rules~\protect\cite{bagan}\\
$0.7\pm 0.1$ & QCD sum rules~\protect\cite{neubertrhosq}\\ 
$0.70 \pm 0.25$ & QCD sum rules~\protect\cite{bs}\\ 
$1.00 \pm 0.02$ & QCD sum rules~\protect\cite{narison}\\ \hline
$0.9\errr{0.2}{0.3}\errr{0.4}{0.2}$&Lattice QCD~\protect\cite{ukqcdiw}
\\ \hline
\end{tabular}
\caption{Values of the Slope of the IW--function of a heavy meson, obtained
using QCD sum rules or Lattice QCD.}
\label{tab:rhosq}
\end{table}

Recently, using analyticity and unitarity properties of the
amplitudes, as well as the heavy quark symmetry, Caprini and Neubert
have derived an intriguing result for the curvature parameter $\hat
c$~\cite{caprini}:
\begin{equation}
\hat c \simeq 0.66\, \hat\rho^2 - 0.11 \ .
\label{eq:chat}\end{equation}
This result implies that one of the two parameters in
(\ref{eq:ftaylor}) can essentially be eliminated, simplifying
considerably the extrapolation to $\omega = 1$. Earlier attempts
to exploit similar methods gave weaker bounds on the parameters.

Finally in this section I consider $B\to D$ semileptonic decays,
which are beginning to be measured experimentally~\cite{artuso}
with good precision. Theoretically the first complication is that
the $1/m_Q$ corrections do not vanish at $\omega = 1$. However, as 
pointed out by Shifman and Voloshin~\cite{sv}, these corrections would
vanish in the limit in which the $b$- and $c$-quarks are degenerate.
This leads to a suppression factor 
\begin{equation}
S = \left(\,\frac{M_B-M_D}{M_B+M_D}\,\right)^2\simeq 0.23
\label{eq:sv}\end{equation}
in the $1/m_Q$ corrections, which reduces their size considerably.
Ligeti, Nir, and Neubert estimate the $1/m_Q$ corrections to be
between approximately $-$1.5\% to +7.5\%~\cite{lnn}. The $1/m_Q^2$
corrections for this decay have not yet been studied systematically.

\subsection{Semileptonic ${\mathbf B\to \rho}$ and ${\mathbf B\to \pi}$ 
Decays}
\label{subsec:vub}

In this subsection I consider the semileptonic decays $B\to\rho$ and
$B\to\pi$. They decays are currently being studied experimentally,
with the goal of extracting the $V_{ub}$ matrix element.

Heavy quark symmetry is less predictive for heavy$\to$light decays
than it is for heavy$\to$heavy ones.  In particular, as we have seen
in the preceding subsection, the normalization condition $\xi(1)=1$
was particularly useful in the extraction of $V_{cb}$. There is no
corresponding normalization condition for heavy$\to$light decays.
Heavy quark symmetry does, however, give useful scaling laws for the
behaviour of the form factors with the mass of the heavy quark at
fixed $\omega$:
\begin{equation}
V,A_2,A_0,f^+  \sim  M^{\frac{1}{2}};\ \ \ \ 
A_1, f^0  \sim   M^{-\frac{1}{2}};\ \ \ \ 
A_3  \sim   M^{\frac{3}{2}}\ .
\label{eq:scaling}\end{equation}
Each of the scaling laws in eq.~(\ref{eq:scaling}) is valid up to
calculable logarithmic corrections.

Several groups have tried to evaluate the form factors using lattice
simulations~\cite{elc}--\cite{ukqcdbtorho} (for a review see
ref.~\cite{jmfstlouis}). The results that I will use for illustration
are taken from the UKQCD collaboration, who have attempted to study
the $q^2$ dependence of the form factors.

From lattice simulations we can only obtain the form factors for part
of the physical phase space.  In order to keep the discretization
errors small, we require that the three-momenta of the $B$,
$\pi$ and $\rho$ mesons be small in lattice units. This implies that we can
only determine the form factors at large values of momentum transfer
$q^2 = (p_B-p_{\pi,\rho})^2$. Fortunately, as we will see below, for
$B\to\rho$ decays, this region of momentum space is appropriate for
the extraction of $V_{ub}$.

\begin{figure}
\begin{picture}(120,200)
\put(50,0){\ewxy{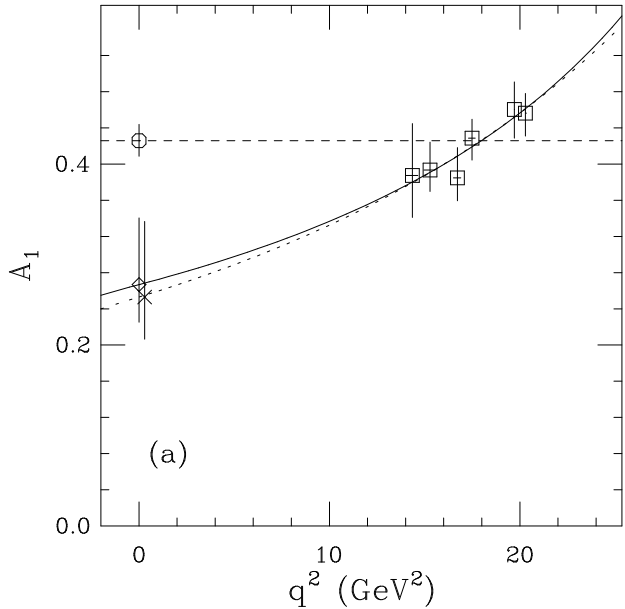}{110mm}} 
\end{picture}
\caption{The form factor $A_1(q^2)$ for the decay
  $\bar B^0\to\rho^+l^-\bar\nu_l$. Squares are measured lattice data,
  extrapolated to the $B$ scale at fixed $\omega$. The three curves
and points at $q^2 = 0$ have been obtained by fitting the squared
using the three procedures described in the text: constant
(dashed line and octagon), pole (solid line and diamond) and
dipole (dotted line and cross).}
\label{fig:ukqcda1}\end{figure}

As an example, I show in fig.~\ref{fig:ukqcda1} the values of the
$A_1$ form factor from ref.~\cite{ukqcdbtorho}.  These authors
evaluate the form factors for four different values of the mass of the
heavy quark (in the region of that of the charm quark), and then
extrapolate them, using the scaling laws in eq.~(\ref{eq:scaling}), to
the $b$-quark.  The squares in fig.~\ref{fig:ukqcda1} represent the
extrapolated values, and as expected they are clustered at large
values of $q^2$.  In order to estimate them over the full kinematical
range some assumption about the $q^2$ behaviour is required.
Fig.~\ref{fig:ukqcda1} also contains three such extrapolations in
$q^2$, performed assuming that:
\begin{itemize}
\item[i)] $A_1$ is independent of $q^2$ (dashed line). The
  extrapolated value of $A_1(0)$ is denoted by an octagon, and the
  $\chi^2/$dof is poor for this fit.
\item[ii)] The behaviour of $A_1(q^2)$ satisfies pole dominance, i.e.
  that $A_1$ is given by
\begin{equation}
A_1(q^2) = \frac{A_1(0)}{(1 - q^2/M_n)^n}\ ,
\label{eq:multipole}\end{equation}
with $n=1$ (solid line). $A_1(0)$ and $M_1$ are parameters of the fit,
but the value of $M_1$ is in the range expected for the $1^+\ b\bar u$
resonance.  The extrapolated value of $A_1(0)$ is denoted by the
diamond.
\item[iii)] The behaviour of $A_1(q^2)$ takes the dipole form
  (\ref{eq:multipole}) with $n=2$ (dotted line). This is almost
  indistinguishable from the pole fit.  The extrapolated value of
  $A_1(0)$ is now denoted by a cross.
\end{itemize}
The $\chi^2/$dof for the pole and dipole fits are both very good. 

\begin{figure}
\begin{picture}(120,200)
\put(50,0){\ewxy{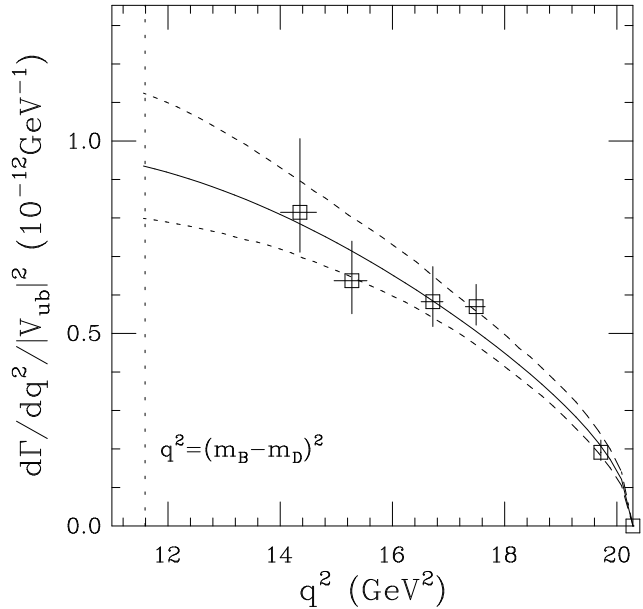}{110mm}} 
\end{picture}
\caption{Differential decay rate as a function of $q^2$ for the semileptonic
  decay $\bar B^0\to\rho^+l^-\bar\nu_l$. Squares are measured lattice
  data, solid curve is fit from eq.~(\protect\ref{eq:distr2}) with 
  parameters given in eqs.~(\protect\ref{eq:asq})
  and (\protect\ref{eq:bsq}). The vertical dotted line marks the charm
  threshold.}
\label{fig:vub}\end{figure}

The UKQCD collaboration~\cite{ukqcdbtorho} comment that for $b\to\rho$
decays in particular, the fact that the lattice results are obtained
at large values of $q^2$ is not a serious handicap to the extraction
of the $V_{ub}$ matrix element. Indeed they advocate using the
experimental data at large values of $q^2$ (as this becomes available
during the next few years) to extract $V_{ub}$. To get some idea of the
precision that might be reached they parametrize the distribution by:
\begin{equation}
\frac{d\Gamma(\bar B^0\to\rho^+l^-\bar\nu)}{dq^2}
 =  10^{-12}\,\frac{G_F^2|V_{ub}|^2}{192\pi^3M_B^3}\,
q^2 \, \lambda^{\frac{1}{2}}(q^2)
 \, a^2\left( 1 + b(q^2 - q^2_{max})\right)\ ,
\label{eq:distr2}\end{equation}
where $a$ and $b$ are parameters to be determined from lattice computations,
and the phase-space factor $\lambda$ is given by $\lambda(q^2)
= (M_B^2+M_\rho^2 - q^2)^2 - 4 m_B^2M_\rho^2$. Already from their current
simulation the UKQCD collaboration are able to obtain $a^2$ with good
precision~\cite{ukqcdbtorho}
\begin{equation}
a^2 = 21\pm 3\ {\mathrm GeV}^2\ .
\label{eq:asq}\end{equation}
Although $b$ is obtained with less precision,
\begin{equation}
b = (-8 \er{4}{6}\,)\,10^{-2}\ {\mathrm GeV}^{-2}\, ,
\label{eq:bsq}\end{equation}
the fits are less sensitive to this parameter at large $q^2$.  The
prediction for the distribution based on these numbers is presented in
fig.~\ref{fig:vub}, and the UKQCD collaboration estimate that they
will be able to determine $V_{ub}$ with a precision of about 10\% or
better.

Although, in this case, the difficulty of extrapolating lattice
results from large values of $q^2$ to smaller ones may not have
significant implications for extracting physical information, this is
not always the case. Already for $B\to\pi$ decays, using results at
large values of $q^2$ restricts the precision with which $V_{ub}$ can
be extracted. This problem is even more severe for the
penguin-mediated rare decay $B\to K^*\gamma$, where the physical
process occurs at $q^2=0$. Much effort is being devoted to this
extrapolation, trying to include the maximum number of constraints
from heavy quark symmetry (as discussed above) and
elsewhere~\cite{extrapolations}. A simple example of such a constraint
for $B\to\pi$ semileptonic decays is that at $q^2=0$, the two form
factors $f^+$ and $f^0$ must be equal. Similar constraints exist for
other processes.

An interesting approach to the problem of the extrapolation to low
values of $q^2$ has been suggested by Lellouch~\cite{lpl}. By
combining lattice results at large values of $q^2$ with kinematical
constraints and general properties of field theory, such as unitarity,
analyticity and crossing, he is able to tighten the bounds on form
factors at all values of $q^2$. This technique can, in principle, be
used with other approaches, such as sum rules, quark models, or even
in direct comparisons with experimental data, to check for
compatibility with QCD and to extend the range of results.

\begin{figure}
\begin{center}
\begin{picture}(180,100)(20,10)
\SetWidth{2}\ArrowLine(10,41)(43,41)
\Text(15,35)[tl]{${\mathbf B}$}
\SetWidth{0.5}
\Oval(100,41)(20,50)(0)
\SetWidth{2}\ArrowLine(157,41)(190,41)
\Text(180,35)[tl]{${\mathbf \rho}$}
\SetWidth{0.5}
\Photon(100,61)(140,101){3}{5}
\Line(140,101)(150,108)\Line(140,101)(150,94)
\Text(158,101)[l]{leptons}
\Text(113,89)[r]{${\mathbf W}$}
\GCirc(50,41){7}{0.8}\GCirc(150,41){7}{0.8}
\Text(75,48)[b]{$b$}\Text(122,48)[b]{$u$}
\Text(100,16)[t]{$\bar q$}
\ArrowLine(101,21)(99,21)
\ArrowLine(70,57)(72,58)\ArrowLine(128,57.5)(130,56.5)
\ZigZag(92,60.5)(92,22){2}{6}
\Text(98,41)[l]{$g$}
\end{picture}
\caption{Representation of a contribution to the semileptonic 
  $B\to\rho$ decay.}
\label{fig:bbkinematics}\end{center}
\end{figure}
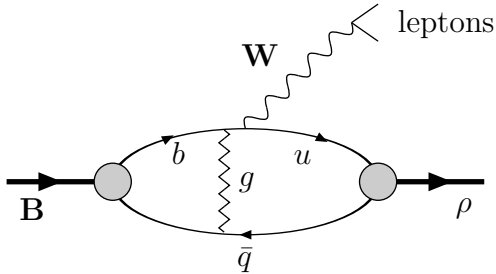

Ball and Braun have recently re-examined $B\to\rho$ decays using
light-cone sum rules~\cite{patricia}, extending the earlier analysis
of ref.~\cite{abs}. Consider, for example, the graph of
fig.~\ref{fig:bbkinematics}, which represents a contribution to the
decay amplitude. For large heavy-quark masses and small $q^2$ there
are two competing contributions of the same order (e.g.
$O(m_Q^{-3/2})$ for the form factor $A_1$). The first one comes from
the region of phase space in which the momentum of the gluon ($g$) is
of the order of $\sqrt{m_b\Lambda_{QCD}}$, so that this contribution
corresponds to small transverse separations and can be treated in
perturbation theory (the non-perturbative effects are contained in the
wave functions at the origin, i.e. in the decay constants). This is
similar to the treatment of hard exclusive processes, such as the form
factors of the pion and the proton at large momentum transfers.
However, there is a second contribution in which the $\rho$-meson is
produced in a very asymmetric configuration with most of the momentum
carried by one of the quarks.  In this case there are no hard
propagators. For most other hard exclusive processes the ``end-point''
contribution is suppressed by a power of the large momentum transfer.
Although, in principle, for $m_Q$ very
large, the end-point is suppressed by Sudakov factors~\cite{asy}, this
suppression is not significant for the $b$-quark.  The end-point
contribution has to be included and treated non-perturbatively, 
since it comes from the region of large transverse separations.
This is the motivation for introducing light-cone sum
rules~\cite{abs}, based on an expansion of operators of increasing
twist (rather than dimension). The non-perturbative effects are
contained in the light-cone wave function of the $\rho$-meson, and the
leading twist contribution to this wave function was recently
re-examined in ref.~\cite{ballbraun}.

An interesting consequence of the analysis of the previous paragraph
is a set of scaling laws for the behaviour of the form factors with
the mass of the heavy quark at fixed (low) $q^2$, rather than at fixed
$\omega$ as in eq.~(\ref{eq:scaling}).  An example of fixed $q^2$
scaling laws is:
\begin{equation}
A_1(0)\,\Theta\,M_P^{3/2} = \mbox{const} (1 + \gamma/M_P + \delta/M_P^2
\ + \cdots)\ ,
\label{eq:scalingm}\end{equation}
where $M_P$ is the mass of the heavy pseudoscalar meson. The factor
$\Theta$ contains the perturbative logarithmic corrections.

\begin{figure}
\begin{picture}(250,250)
\put(-20,20){\ewxy{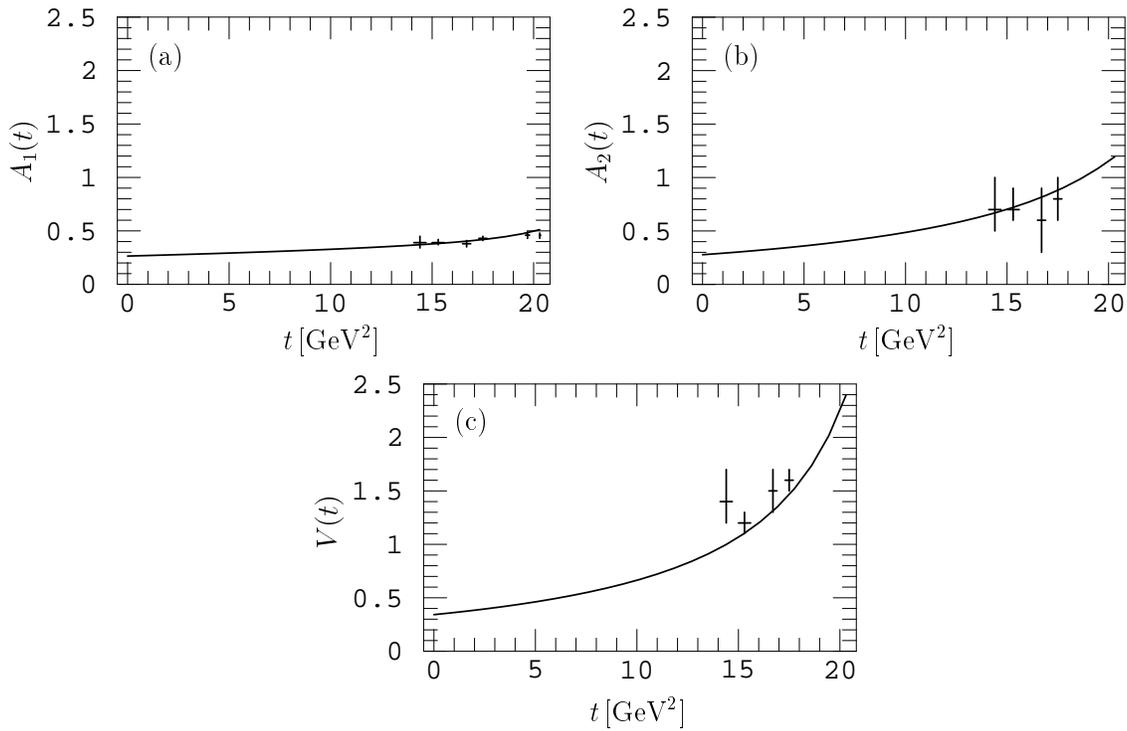}{200mm}}
\end{picture}
\caption{Results for the form factors $A_1(q^2)$, $A_2(q^2)$ and 
  $V(q^2)$ for $B\to\rho$ semileptonic decays as a function of
  $t=q^2$~\protect\cite{patricia}. The curves correspond to the
  results obtained with light-cone sum rules by Ball and
  Braun~\protect\cite{patricia}, and the points to the results from
  the UKQCD collaboration~\protect\cite{ukqcdbtorho}.}
\label{fig:pball}\end{figure}

Some of the results of Ball and Braun are presented in
fig.~\ref{fig:pball}, where the form factors $A_1, A_2$ and $V$ are
plotted as functions of $q^2$. The results are in remarkable agreement
with those from the UKQCD collaboration, in the large $q^2$ region
where they can be compared.

\section{Conclusions}
\label{sec:concs}

The principal difficulty in deducing weak interaction properties from
experimental measurements of $B$-decays lies in controlling the strong
interaction effects. These are being studied using non-perturbative
methods such as lattice simulations or QCD sum rules. Considerable
effort and progress is being made in reducing the systematic
uncertainties present in lattice computations.

Although both the theoretical and experimental errors on the value of
$f_{D_s}$ are still sizeable, it is nevertheless very pleasing that
they are in agreement. It is also satisfying that the values of
$V_{cb}$ extracted from exclusive and inclusive measurements are in
good agreement.  The theoretical uncertainties for the two processes
are different, and the agreement is evidence that they are not
significantly underestimated.

It has been argued that $B\to\rho$ decays at large $q^2$, where the
evaluation of the relevant form factors using lattice simulations is
reliable, will soon provide a determination of $V_{ub}$ at the 10\%
level or better~\cite{ukqcdbtorho}. It will also be interesting to
observe developments of the light-cone approach to these decays.

Many lattice computations of $f_B$ have been performed using static
heavy quarks ($m_Q=\infty$), and serve as a very valuable check of the
consistency of the extrapolation of the results obtained with finite
heavy-quark masses. Such checks have not been performed yet for many
other quantities in $B$-physics; this is an important omission, which
should be put right.

This talk has been about the decays of $B$-mesons. Detailed
experimental and theoretical studies are also beginning for the
$\Lambda_b$-baryon. For example, the first lattice results for the
Isgur--Wise function of the $\Lambda_b$ has been presented in
ref.~\cite{lambdab}.

\subsection*{Acknowledgements}
It is a pleasure to thank Nando Ferroni and the other organizers of
Beauty `96 for the opportunity to participate in such an interesting
and stimulating meeting. I gratefully acknowledge many helpful and
instructive discussions with Patricia Ball, Volodya Braun, Jonathan
Flynn, Laurent Lellouch, Guido Martinelli, Matthias Neubert, Juan
Nieves, Nicoletta Stella and Kolya Uraltsev. I also acknowledge the
Particle Physics and Astronomy Research Council for their support
through the award of a Senior Fellowship.


\begin{thebibliography}{99}
  
\bibitem{kolya} N.G.~Uraltsev, these proceedings. 
\bibitem{ahmed} A.~Ali, these proceedings.
\bibitem{gronau} M.~Gronau, these proceedings.
\bibitem{mn} M.~Neubert Phys. Rep. \underline{245} (194) 259.
\bibitem{ms} M.A.~Shifman, hep-ph/9510377, Lecture given at the Theoretical
Advanced Study Institute, {\em QCD and Beyond}, Boulder, June 1995.
\bibitem{lattbphys} H.~Wittig, hep-ph/9606371, to be published in the
proceedings of the 3rd German--Russian Workshop on Progress in Heavy
Quark Physics, Dubna, Russia, 20-22 May 1996.
\bibitem{perfect}  P. Hasenfratz and F. Niedermayer,
Nucl.~Phys.~\underline{B414} (1994) 785.
\bibitem{improvement} K.~Symanzik, in ``Mathematical Problems in Theoretical
Physics'', \\ Springer Lecture Notes in Physics,
vol. 153 (1982) 47, ed. R.~Schrader, R.~Seiler and D.A.~Uhlenbrock.
\bibitem{svz} M.A.~Shifman, A.I.~Vainshtein and V.I.~Zakharov,
Nucl.~Phys.~\underline{B147} (1979) 385 and 448.
\bibitem{jlqcdbb} JLQCD Collaboration, S.~Aoki et al., 
Nucl. Phys. \underline{B (Proc.Suppl) 47} (1996) 433
\bibitem{bbs} C.~Bernard, T.~Blum and A.~Soni, hep-lat/9609005 (1996) 
\bibitem{marseille} C.T.~Sachrajda, Proceedings of the 1993 EPS
Conference on High Energy Physics, Marseille, France, July 1993
(eds J.~Carr and M.~Perrottet, Editions Fronti\`eres, Gif-sur-Yvette, 1994)
p. 957
\bibitem{allton} C.R.~Allton, Nucl.~Phys.~\underline{B\,(Proc.Suppl.)\,47}
(1996) 31.
\bibitem{milc} MILC Collaboration, C.~Bernard et al., hep-lat/9608092
\bibitem{jlqcdfds} JLQCD Collaboration, A.~Aoki et al., hep-lat/9608142
\bibitem{pdg} R.M.~Barnett et al., Phys.~Rev.~\underline{D54} (1996) 1
\bibitem{cleoupdate} CLEO collaboration, D.~Gibaut et al., 
CLEO CONF 95-22 (1995)
\bibitem{e653} Fermilab E653 Collaboration, K.~Kodama et al., hep-ex/9606017
\bibitem{richman} J.~Richman, to be published in the proceedings of the 
1996 ICHEP Conference, Warsaw, July 1996.
\bibitem{narisoncracow} S.~Narison, Acta~Phys.~Polon.~\underline{B26}
(1995) 687.
\bibitem{luke} M.~Luke, Phys.~Lett.~\underline{B252} (1990) 447
\bibitem{czarnecki} A.~Czarnecki, Phys. Rev. Lett. \underline{76} (1996)
4124
\bibitem{fn} A.~Falk and M.~Neubert, Phys.~Rev.~\underline{D47} (1993)
2965 and  2982.
\bibitem{mannel} T.~Mannel. Phys.~Rev.~\underline{D50} (1994) 428.
\bibitem{suv} M.A.~Shifman, N.G.~Uraltsev and A.I.~Vainshtein, 
Phys.~Rev.~\underline{D51} (1995) 2217.
\bibitem{ht} G.~Martinelli and C.T.~Sachrajda, 
hep-ph/9605336 (1996).
\bibitem{artuso} M.~Artuso, these proceedings.
\bibitem{neubertcernschool} M.~Neubert, 
Int.~J.~Mod.~Phys.~\underline{A11} (1996) 4173.
\bibitem{bj} J.D.~Bjorken in {\em Results and Perspectives in Particle
Physics}, proceedings of the 4th Rencontres de Physique de la Vall\'ee
d'Aoste, La Thuile, 1990, edited by M.~Greco (Editions Fronti\`eres,
Gif-sur-Yvette, 1990) p. 583; in 
{\em Gauge Bosons and Heavy Quarks}, proceedings
of the 18th SLAC Summer Institute on Particle Physics, Stanford, California,
1990, edited by J.F.~Hawthorne, SLAC Report 378 (1991) p. 167. 
\bibitem{voloshin} M.B.~Voloshin, Phys.~Rev.~\underline{D46} (1992) 3062.  
\bibitem{gk} A.~Grozin and G.~Korchemsky, reported in
    ref.~\cite{neubertcernschool}.
\bibitem{kn} G.~Korchemsky and M.~Neubert, reported in
    ref.~\cite{neubertcernschool}.
\bibitem{bagan} E.~Bagan, P.~Ball and P.~Gosdzinsky, 
Phys.~Lett.~\underline{B301} (1993) 249.
\bibitem{neubertrhosq} M.~Neubert, Phys.~Rev.~\underline{D47} (1993) 4063.
\bibitem{bs} B.~Blok and M.A.~Shifman, Phys.~Rev.~\underline{D47} (1993) 2949.
\bibitem{narison} S.~Narison, Phys.~Lett.~\underline{B325} (1994) 197.
\bibitem{ukqcdiw} UKQCD Collaboration, K.C.~Bowler et al., 
Phys.~Rev.~\underline{D52} (1995) 5067 
\bibitem{cleoii} CLEO Collaboration, B.~Barish et al., 
Phys.~Rev.~\underline{D51} (1995) 1014
\bibitem{caprini} I.~Caprini and M.~Neubert,
 Phys.~Lett.~\underline{B380} (1996) 376.
\bibitem{sv} M.B.~Voloshin and M.A.~Shifman, Yad.~Fiz.~\underline{45} (1987)
463 and \underline{47} (1987) 801 [Sov.~J.~Nucl.~Phys.~\underline{45} (1987) 
292 and \underline{47} (1988) 511].
\bibitem{lnn} Z.~Ligeti, Y.~Nir and M.~Neubert, 
Phys.~Rev.~\underline{D49} (1994) 1302.
\bibitem{elc} As.Abada et al., Nucl.~Phys.~\underline{B416} (1994) 675.
\bibitem{apebtorho} APE Collaboration, C.R. Allton et al.,
Phys.~Lett.~\underline{B345} (1995) 513.
\bibitem{ukqcdbtorho} UKQCD Collaboration, J.M.~Flynn et al., 
Nucl. Phys. \underline{B461} (1996) 327
\bibitem{jmfstlouis} J.M.~Flynn, to be published in the proceedings of the
1996 International Conference on Lattice Field Theory, St.Louis, June 1996.
\bibitem{extrapolations} UKQCD Collaboration, J.M. Flynn et al.,
hep-ph/9602201 (1996).
\bibitem{lpl} L.P.~Lellouch, hep-ph/9509358 (1995)
\bibitem{patricia} P.~Ball, hep-ph/9605233; V.M.~Braun, hep-ph/9510404 (1995);
P.~Ball and V.M.~Braun, in preparation.
\bibitem{abs} A.~Ali, V.~Braun and H.~Simma, Z.~Phys.~\underline{C63} (1994) 
437.
\bibitem{asy} R.~Akhoury, G.~Sterman and Y.P.~Yao,
Phys.~Rev.~\underline{D50} (1994) 358.
\bibitem{ballbraun} P.~Ball and V.M.~Braun, Phys.Rev. \underline{D54} (1996) 
2182.
\bibitem{lambdab} N.~Stella (for the UKQCD Collaboration), hep-lat/9607072 
(1996).
\end{thebibliography}
\end{document}